# Benchmark Study of Transient Stability during Power-Hardware-in-the-Loop and Fault-Ride-Through capabilities of PV inverters

Carina Lehmal, Ziqian Zhang, Robert Schürhuber
Institute of Electrical Power Systems
Graz University of Technology
Graz, Austria
ziqian.zhang@tugraz.at

*Abstract—* The deployment of PV inverters is rapidly expanding across Europe, where these devices must increasingly comply with stringent grid requirements. This study presents a benchmark analysis of four PV inverter manufacturers, focusing on their Fault Ride-Through capabilities under varying grid strengths, voltage dips, and fault durations—parameters critical for grid operators during fault conditions. The findings highlight the influence of different inverter controls on key metrics such as total harmonic distortion of current and voltage signals, as well as system stability following grid faults. Additionally, the study evaluates transient stability using two distinct testing approaches. The first approach employs the current standard method: testing with an ideal voltage source. The second utilizes a Power-Hardware-in-the-Loop methodology with a benchmark CIGRE grid model. The results reveal that while testing with an ideal voltage source is cost-effective and convenient in the short term, it lacks the ability to capture the dynamic interactions and feedback loops of physical grid components. This limitation can obscure critical real-world factors, potentially leading to unexpected inverter behavior and operational challenges in grids with high PV penetration. This study underscores the importance of re-evaluating conventional testing methods and incorporating Power-Hardware-in-the-Loop structures to achieve test results that more closely align with real-world conditions.

***Benchmark-Studie der transienten Stabilität und FRT Funktionalität von PV Wechselrichtern***

Die Installation von PV-Wechselrichtern nimmt in ganz Europa seit der letzten Jahre rapide zu, wobei die Wechselrichter zunehmend strengeren Netzanforderungen entsprechen müssen. Diese Studie präsentiert eine Benchmark-Analyse von vier PV-Wechselrichten bekannter Hersteller mit einem Schwerpunkt auf ihren Fault-Ride-Through Funktionalitäten bei unterschiedlichen Netzstärken, Spannungseinbrüchen und Fehlerdauern – Parameter, die für Netzbetreiber in Fehlerfällen von entscheidender Bedeutung sind. Die Ergebnisse zeigen den Einfluss verschiedener Software-Implementierungen auf zentrale Kennwerte wie die Gesamtklirrfaktoren von Strom- und Spannungsverläufe sowie die Stabilität nach Netzfehlern. Darüber hinaus wurde die transiente Stabilität mit zwei unterschiedlichen Testansätzen untersucht. Der erste Ansatz verwendet die derzeitige Standardmethode: Kompatibilität beim Einsatz einer idealen Spannungsquelle. Der zweite Ansatz nutzt eine Power-Hardware-in-the-Loop Methodik mit einem CIGRE-Benchmark-Netzmodell im Hintergrund. Die Ergebnisse verdeutlichen, dass Tests mit einer idealen Spannungsquelle zwar kurzfristig kosteneffizient und praktisch sind, jedoch nicht in der Lage sind, die dynamischen Wechselwirkungen und Rückkopplungen physischer Netzkomponenten vollständig abzubilden. Diese Einschränkung kann entscheidende Faktoren verschleiern und zu unerwartetem Verhalten der Wechselrichter sowie zu betrieblichen Herausforderungen in Netzen mit hoher PV-Durchdringung führen. Die Studie unterstreicht die Bedeutung einer Neubewertung konventioneller Testmethoden und die Integration von Power-Hardware-in-the-Loop-Strukturen, um Testergebnisse zu erzielen, die näher an realen Bedingungen liegen.

*Schlüsselwörter: Transiente Stabilität, Laboruntersuchung, Power-Hardware-in-the-Loop Test, FRT-Funktionalität, PV Wechselrichter*

## I. INTRODUCTION

With the focus on the energy transition, photovoltaic (PV) power generation has emerged as a critical player due to its geographic adaptability across Europe. The decreasing costs of PV modules and inverters have made PV generation an efficient solution in terms of land use, adaptable for rooftop installations, and economically viable. Consequently, power generated by PV inverters now plays a vital role within Europe's ENTSO-E grid system.

The quality of PV-generated power—specifically total harmonic distortion (THD) and stability following grid faults—has significant implications for overall grid performance. Increasingly, partial brownouts and blackouts caused by inverter oscillations highlight the risks of widespread inverter use. Notable examples include the 2016 blackout in Australia [1] and the 2019 blackout in the UK [2]. These incidents demonstrate that components within inverters, such as power electronics and control software, and their dynamic interactions with the grid can lead to instability. As the share of renewable energy grows, precise evaluation of inverters during compatibility testing is critical to maintaining grid stability.

In many countries, compliance with grid standards is a prerequisite for PV inverter installation. Accredited testing institutes evaluate inverters against established testing standards, ensuring their behavior during grid faults poses no risk to the grid. This process supports grid operators in making accurate forecasts and maintaining voltage and angle stability to prevent outages.

In Austria, PV inverters are highly regulated, and only those listed on the Whitelist published by Oesterreich Energie [3] are permitted for installation. Certification involves analyzing both hardware and software, including active and reactive power control, to ensure compliance with national grid standards (e.g., TOR Type A). Since inverter performance heavily depends on internal software parameters and control loops [4], conventional physical generator equations no longer reliably describe their behavior. As control time constants also influence stability and are often not externally adjustable, compatibility testing is essential for integrating inverters into increasingly decentralized, flexible, and volatile grids.

## II. TESTING METHODS

To evaluate PV inverters, it is essential to understand their operational role within the grid. The grid comprises an infinite voltage source, transmission lines, transformers, and inverters. Inverters monitor terminal voltage to extract grid condition data—amplitude, phase, and frequency—while control algorithms convert DC power from PV modules into AC power synchronized with grid voltage.

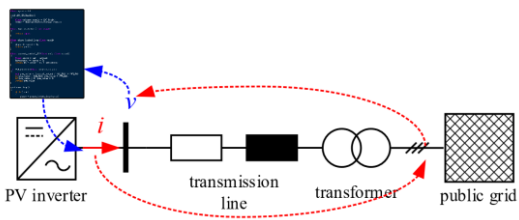

*Figure 2.1: Feedback Relationship Diagram of the Photovoltaic Inverter-Grid System*

However, complexities arise when there is impedance between the inverter and the voltage source, typically due to cables and transformers as seen in Figure 1. This impedance creates a feedback loop where the inverter's current output can affect terminal voltage. Inadequate control algorithms may struggle with voltage fluctuations, leading to instability and, in severe cases, grid issues such as islanding [5-7].

During short-circuit faults, the fault impedance impacts the inverter, which must adapt to maintain stability. Advanced control strategies are required to adjust output current in response to terminal voltage changes. After fault clearance, inverters must swiftly stabilize to prevent grid disturbances. The dynamic demands on inverters underscore the necessity for rigorous testing methods.

*2.1 Testing with an Ideal Voltage Source*

The ideal voltage source method is widely used to test PV inverters. Certification labs simulate grid conditions using a three-phase sinusoidal voltage source capable of programmed voltage drops. This source maintains a pre-defined voltage waveform regardless of load, emulating a grid with zero internal resistance (Figure 2.2).

However, this approach has limitations. By ignoring feedback interactions between the inverter and the grid, it simplifies testing conditions and reduces the standards required of inverters. While inverters may perform well in this setup, their fault response and recovery capabilities under real-world conditions remain unverified. Consequently, relying solely on this method risks overestimating inverter performance, posing potential threats to grid stability.

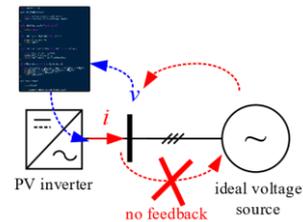

*Figure 2.2: Photovoltaic Inverter - Ideal Voltage Source Schematic*

*2.2 Testing wit PHIL Systems*

Simulating real-world grid conditions, including short-circuit faults, in laboratory settings is complex and often impractical. Physical grid setups introduce risks to external equipment and lack flexibility, as even minor scenario changes require significant reconfiguration.

Power Hardware-in-the-Loop (PHIL) systems address these challenges [8]. Similar to a flight simulator used for pilot training, PHIL systems create a "simulated cockpit" for testing electrical equipment. Using real-time simulators and power amplifiers, PHIL systems replicate grid conditions, including dynamic interactions and feedback loops. In PHIL testing, the real-time simulator generates electrical signals that represent grid conditions. These signals are converted into voltage or current inputs for the inverter under test. Simultaneously, the inverter's output signals are fed back into the simulator to update the grid model state. This closed-loop process evaluates the inverter's stability and fault response under realistic operational scenarios.

Compared to ideal voltage source tests, PHIL systems provide a more realistic environment, enabling a comprehensive evaluation of inverter performance under real-world conditions. The software-based grid model in PHIL setups is highly flexible, allowing easy adjustments to testing parameters via a computer interface. PHIL technology significantly enhances testing accuracy and ensures that PV inverters meet the high dynamic requirements necessary for maintaining grid stability and safety in decentralized energy systems.

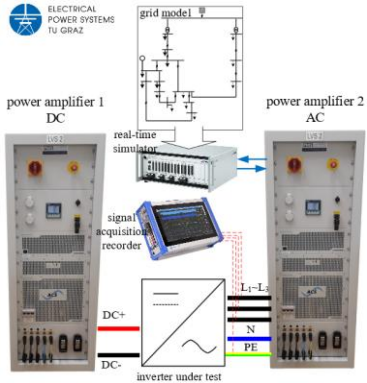

*Figure 2.3: PHIL test schematic*

All four inverters were mounted on a wooden panel for ease of installation and accessibility. The laboratory setup is shown in the figure below:

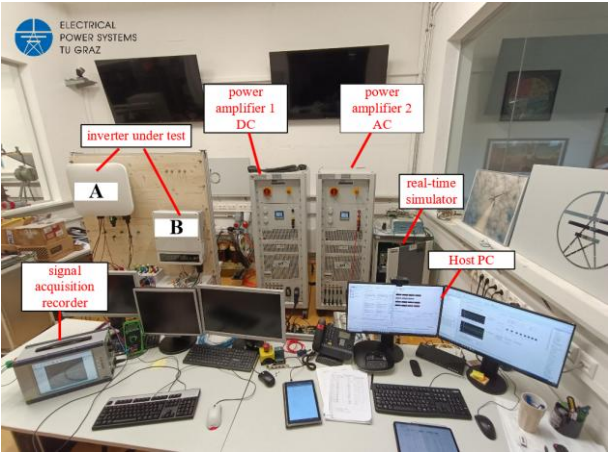

*Figure 2.4: PHIL test laboratory*

### III. BENCHMARK STUDY OF FRT BEHAVIOUR

For this benchmark study, four models of 10 kW inverters with similar operating voltage ranges are analyzed. The study provides an in-depth assessment of the transient stability performance of PV inverters, focusing on their dynamic response under weak grid conditions and performance during AC grid faults. Key aspects of evaluation included inverter stability during faults, FRT capability, and recovery performance after fault clearance. The testing framework was based on the Austrian guideline TOR-Type A [9], ensuring a thorough and impartial evaluation.

When grid voltage drops occur, PV inverters are expected to maintain normal operation unless the voltage level and fault duration exceed the limits defined in the grid code. These limits are specified by an envelope curve, as shown in Figure 3.2. According to the grid code, an inverter may only disconnect from the grid when faults surpass the envelope curve's range (represented by the green area outside the red zone in Figure 3.1). If the fault remains within the envelope curve's defined range (the red zone in Figure 3.1), the inverter must recover quickly, returning to its pre-fault active current output within one second of fault clearing.

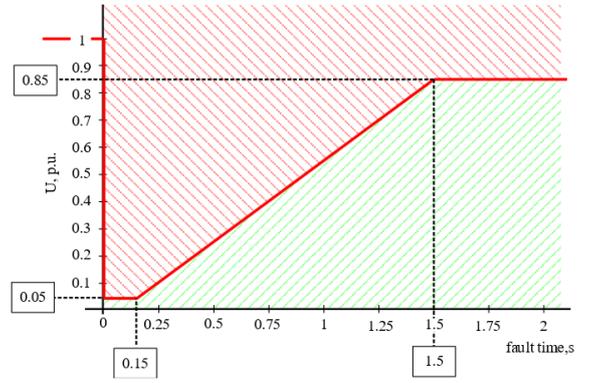

*Figure 3.1: LVRT Curve according TOR Type A*

A PV inverter capable of providing stable and sinusoidal short-circuit current during faults supports accurate fault detection by protection devices. This capability enables rapid disconnection of the faulted line, ensuring effective fault clearance and safeguarding equipment and personnel. For voltage stability in medium voltage networks, PV inverters must also deliver stable reactive current during faults to maintain adequate voltage levels. This ensures that the voltage remains relatively high, enabling the normal operation of other grid-connected equipment and preserving overall grid stability.

For the benchmark study, the ideal voltage source method was first analyzed as a baseline testing approach. This was then compared to the PHIL testing model, which incorporated the medium-voltage grid model developed by the CIGRE Working Group C6.04.02 [10]. Each test was conducted at least three times, varying the fault initiation angle to capture potential differences in inverter response behavior.

To ensure an objective comparison among the four inverters, pre-test configurations are standardized. This included harmonizing grid code settings and threshold values across all devices, accounting for slight variations in their default ride-through settings [6].

Table 1: Key technical parameters of the four inverters:

|   |   | **A** | **B** | **C** | **D** |
|---|---|---|---|---|---|
| **DC** | **Rated input power** | 10 kWp | 10 kWp | 10 kWp | 10 kWp |
|   | **Operation voltage range** | 140-980 V | 200-850V | 160-1000V | 180–950 V |
|   | **Rated input voltage** | 600 V | 620 V | 600 V | 720 V |
| **AC** | **Rated output power** | 10 kW | 10 kW | 10 kW | 10 kW |
|   | **Rated output voltage** | 380/400 $V_{AC}$, 3L+N+PE | | | |
|   | **Rated AC grid frequency** | 50/60 Hz | | | |
|   | **Max output current** | 16.9 A | 16.5 A | 16.7 A | 16 A |

## 3.1 Ideal voltage source testing

When testing with an ideal voltage source, only the amplitude of the AC grid voltage changes during the test, while the phase remains constant. As a result, the phase synchronization control algorithm of the photovoltaic inverters is neither challenged nor disturbed. This makes the ideal voltage source test the least demanding for assessing the transient stability of photovoltaic inverters, as it minimizes the feedback effect of the inverter's output current on the grid voltage. Consequently, all photovoltaic inverters are expected to pass this test without difficulty.

Figure 3.2 illustrates an example of the waveforms from the ideal voltage source, showing a voltage drop to 0.5 p.u for 120ms.

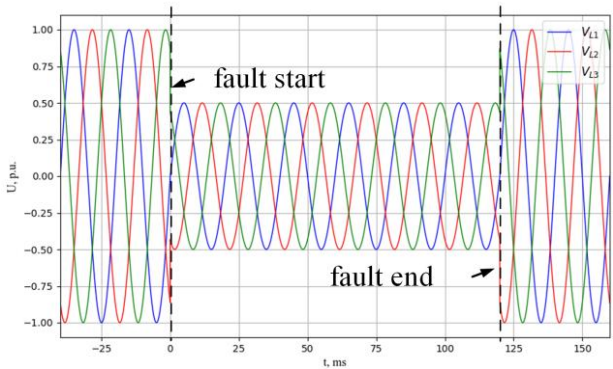

*Figure 3.2: Voltage curve example during ideal voltage source test*

## 3.2 PHIL System testing

To more accurately assess the transient stability of PV inverters under real-world conditions, the complex grid model developed by the CIGRE working group was selected. This test case is designed to replicate fault sequences and disturbances typically encountered in grid operations, including the dynamic effects of the protection system. The model provides a comprehensive analysis of the integration of Distributed Energy Resources (DERs) in distribution grids, offering deeper insights.

The PV inverter under test is connected to this grid model via a 20/0.4 kV transformer, as indicated by the red dot in Figure 3.6. Short-circuit fault locations are marked with two lightning symbols, corresponding to test cases 1 and 2. In a closed-ring grid, protection devices at both ends of a faulty line typically trigger circuit breakers to isolate the fault and restore voltage to the unaffected grid.

Due to varying response times of the protection systems, one end of the line disconnects before the other, creating a two-stage residual voltage. Additionally, when the faulty line is disconnected at both ends, the grid topology changes, resulting in a shift in the voltage phase angle. This dynamic fault behavior offers a more realistic and relevant framework for evaluating the performance of PV inverters.

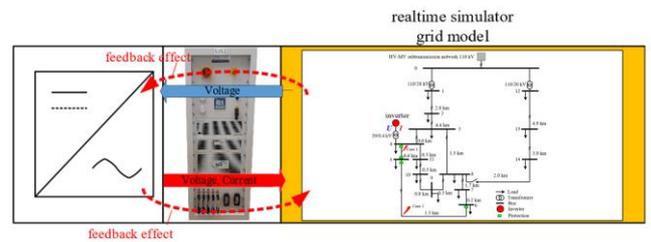

*Figure 3.3: Schematic of PHIL Testing*

## 3.3 Initial Fault Dynamics

### A. Desired condition

At fault initiation the current of the inverter should rise for only such a long time that fault detection schemes can react and accordingly forward their trigger signals which is depending on the device at least 0.4 s. The fault current from the inverter, however, should not rise above levels that are hazardous for the inverter themselves.

Figure 3.4 presents the measured current response of photovoltaic inverters during a fault event. This figure overlays the results from over 100 different test cases, showing the short-circuit current. On the left side, the short-circuit current's probability of occurrence at various time points is represented using a grayscale gradient: the darker areas indicate a higher likelihood of the short-circuit current waveform occurring at that specific time, while the lighter areas correspond to a lower probability. On the right side, the curves are illustrated in color. This visualization allows for a clear identification of the short-circuit current characteristics of the four types of inverters in a single figure. The x-axis represents time in milliseconds, while the y-axis shows the current in per unit (p.u.). The fault begins at 0 ms and is cleared at 120 ms, as specified by the FRT curve.

### B. Assessment

At fault initiation (0 ms), all four inverters experienced an instantaneous current spike exceeding 2 p.u., requiring rapid limitation to protect components. Inverters A and D quickly controlled the surge; however, inverter A sustained a fault current of 1.5 p.u. for 10 ms, while inverter D nearly eliminated it, potentially hindering fault detection. Inverter B failed to reduce its current below 2 p.u. within 20 ms, showing instability. Inverter C reduced its current to 0 p.u., but this also impaired fault identification and displayed some instability.

In the grayscale curves of Figure 3.4, significant differences in the behavior of inverter currents following the initiation of a fault are evident. While Inverters A and B generally maintain a continuous current flow, closely matching the pre-fault current values, Inverters C and D exhibit a distinct response. Both Inverter C and D restart their current at zero and gradually increase it over time. Notably, Inverter C requires approximately one-third of the fault duration to restore its current to pre-fault levels. In contrast, Inverter D often requires a duration longer than the entire fault period to achieve the same current level. This delayed response, particularly in the case of Inverter D, underscores significant differences in performance and fault recovery behavior across the inverters.

*3.4 Post Fault Dynamics*

A. Desired condition

Quickly stabilizing active current post-fault to pre-fault levels is critical for grid health. At best no oscillations should be introduced and the output current should not drop below 1 p.u.

Figure 3.4 also provides direct insights into post-fault dynamics, illustrating the behavior of photovoltaic inverters following fault clearance at 120 ms.

B. Assessment

Inverter A demonstrated excellent stability, with its output current quickly returning to the pre-fault state after the fault was cleared. In contrast, inverter B experienced significant oscillations before eventually stabilizing after approximately 550 milliseconds.

Inverter C recovered to 1.0 p.u. within 250 milliseconds, but its output displayed high harmonic content and occasional spikes, indicating suboptimal performance. Inverter D exhibited pronounced instability, with its current oscillating between 0 and over 2.0 p.u. in a continuous loop until the inverter was either automatically or manually shut down.

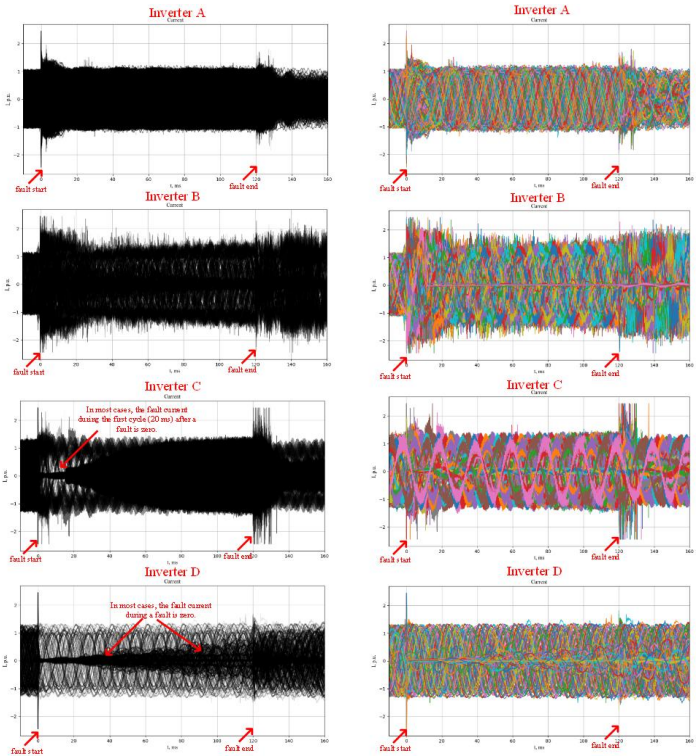

*Figure 3.4: Overlapping waveforms of inverters of ideal voltage source test; (left) curves in grayscale, (right) curves in color*

*3.5 THD assessment*

For the Total Harmonic Distortion (THD) assessment, the current waveforms are subjected to a Fast Fourier Transform (FFT) analysis to obtain their frequency spectra. This detailed examination focused on identifying the frequencies at which spikes occurred, as these provide valuable insights into the internal control parameters and tuning of the inverter's software.

C. THD assessment under ideal voltage source conditions

Inverter A performed exceptionally well, with a total harmonic distortion (THD) of only 2.89%, and was able to stably inject reactive power into the grid during faults, contributing to grid stability. In contrast, inverter B exhibited a THD of 7.03% and did not perform well during grid faults. Inverter C showed an even higher THD of 12.39%, and although it could inject reactive power in some fault conditions, it failed to consistently meet grid codes. Inverter D had the lowest THD at 2.55%, but showed instability after fault clearing.

Overall, under the ideal voltage source test, Inverters A and D demonstrated strong performance in terms of harmonic distortion, but only Inverter A achieved satisfactory fault response and stability.

D. THD assessment under PHIL system conditions

In the PHIL system tests, both fault occurrences and grid strength at the interface are varied. Using the medium-voltage distribution grid model developed by the CIGRE C6.04.02 working group, the performance of the PV inverters differed significantly. Specific protection devices and response times are configured to replicate real-world grid fault complexities.

Inverter A performed excellently, consistently supplying reactive power and quickly resuming normal operation after fault clearing. In contrast, inverters B and C struggled, unable to ride through the faults, with unstable current and power outputs. Inverter D also showed poor performance, failing to inject sufficient reactive power into the grid.

*3.6 Comparison of ideal voltage source and PHIL system behavior*

Performance testing of various photovoltaic inverters revealed notable differences between results obtained using ideal voltage sources and PHIL testing. As illustrated in Figure 3.4, current waveforms in ideal voltage source tests are generally smoother and more stable.

Conversely, PHIL testing exposed issues not evident in ideal voltage source tests. Significant amplitude spikes as well as instability are demonstrated, which contrasts the more stable performance during ideal voltage source tests.

Inverter A, performed consistently well in both scenarios, underscoring its superior transient stability. Inverter C experienced significant oscillations and amplitude spikes, while inverter D displayed marked instability in PHIL conditions, contrasting its more stable performance in ideal voltage source testing.

These findings emphasize that relying solely on ideal voltage source testing risks overly optimistic evaluations, as it does not replicate the complex feedback interactions of real grid conditions. PHIL testing, though more demanding, offers a rigorous evaluation aligned with actual grid dynamics, better predicting inverters' real-world performance.

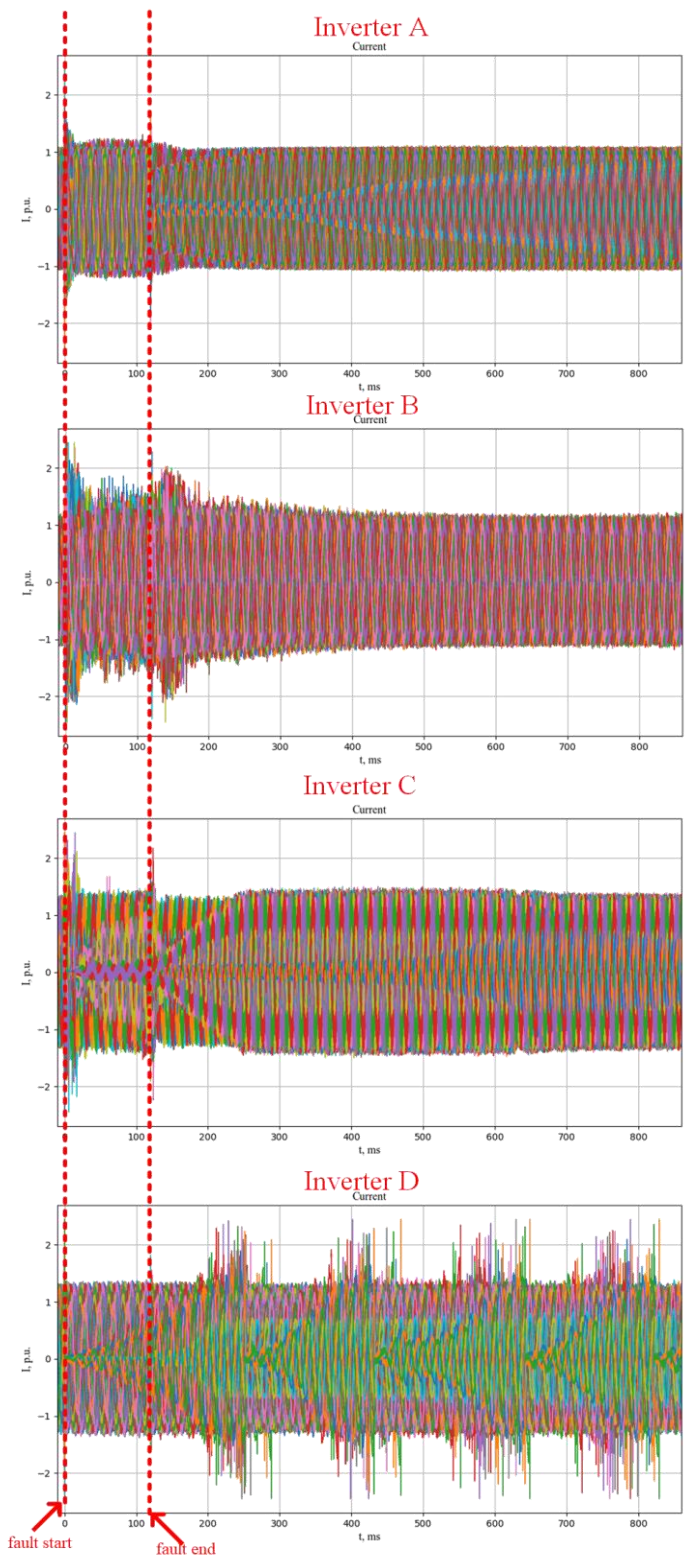
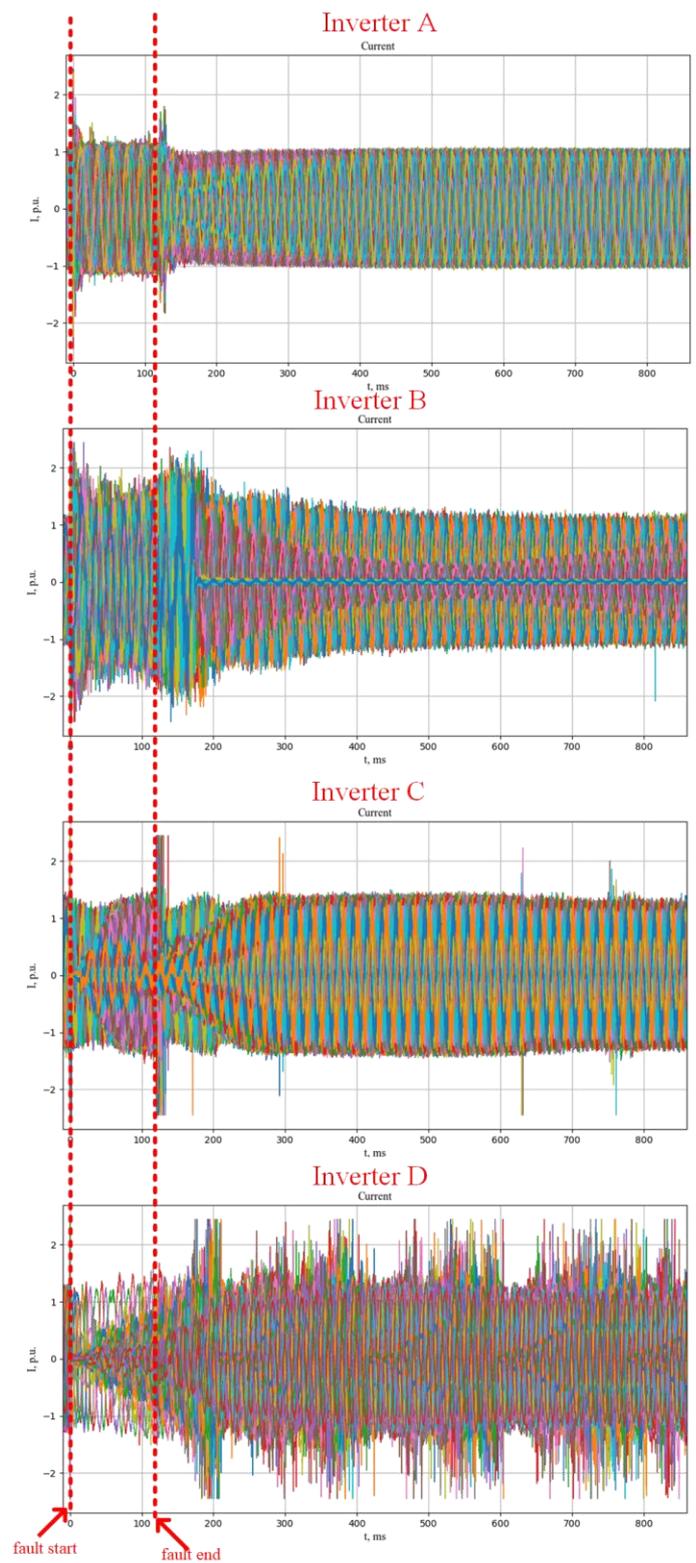

*Figure 3.5: Comparison of overlapping current from ideal voltage source test and PHIL test at high grid strength*

## IV. Conclusion

The benchmark study revealed two key findings. First, the performance and THD values varied significantly depending on the manufacturer. This difference was attributed to variations in the software control loops and the design of the filters at the inverter's interface to the grid. Second, a significant disparity was observed between the results of the ideal voltage source test and the PHIL system test.

The ideal voltage source test lacks the necessary feedback loop between the grid and the inverter, which is crucial for accurately simulating real-world conditions. As a result, it did not reflect the actual behavior of the inverter under real grid conditions. In contrast, the PHIL test, which better captures the dynamic interaction between the inverter and the grid, showed more instability, particularly under weak grid conditions.

The pass rates for the tests are as follows:

- Inverter A: 97.7%

  Inverter A demonstrated strong performance across all test conditions, maintaining stability even at a low grid strength of 2 and achieving a very low THD of 2.89%.

- Inverter B: 23.3%

  Inverter B struggled with stability, especially following fault clearing, and exhibited a higher THD of 7.03%. It only operated stably at a grid strength of 10.

- Inverter C: 48.8%

  Inverter C showed acceptable performance, but it struggled to inject sufficient reactive power during voltage drops and became unstable in weak grid conditions. Its THD was notably high at 12.39%, though it could operate at a grid strength as low as 3.

- Inverter D: 11.6%

  Inverter D performed the poorest. It failed to inject sufficient reactive power in most fault cases and became unstable after fault clearing in most test cases. Despite its lowest THD of 2.55%, its overall performance was unsatisfactory.

Inverter testing is essential for ensuring system reliability, safety, and effective performance evaluation, especially during faults. While the ideal voltage source method is practical and fast, it offers limited insights into inverter-grid interactions and may fail to predict real-world challenges. This method showed a high pass rate of 77.7% for the four tested inverters, largely due to its stable and simplified testing conditions.

In contrast, the PHIL testing method provides a more realistic simulation of grid conditions, exposing weaknesses in device performance.

The pass rate for PHIL testing was notably lower at 36.7%, reflecting its ability to reveal vulnerabilities stemming from grid feedback complexities. A comparison of test results (Figure 3.5) shows that while the ideal voltage source tests indicated stable inverter performance, PHIL testing uncovered significant instabilities, particularly in response to dynamic grid conditions.

Although the ideal voltage source approach is cost-effective and convenient, its inability to simulate realistic fault scenarios can lead to undetected risks. PHIL testing, on the other hand, delivers valuable insights into inverter behavior in actual grid environments, aiding system optimization and ensuring reliability. For practical applications, PHIL testing is strongly recommended to identify potential issues early and ensure stable grid integration.

## V. References

[1] Operator, A. E. M. Integrated Final Report SA Black System 28 September 2016. Australia Energy Market Operator2017, Available: http://www.aemo.com.au/media/Files/Electricity/NEM/Market_Notices_and_Events/Power_System_Incident_Reports/2017/Integrated-Final-Report-SA-Black-System-28-September-2016. pdf.

[2] Bialek, Janusz. "What does the GB power outage on 9 August 2019 tell us about the current state of decarbonised power systems?." Energy Policy 146 (2020): 111821.

[3] https://oesterreichsenergie.at/publikationen/ueberblick/detailseite/wechselrichterliste-tor-erzeuger-typ-a

[4] Lehmal, Carina, et al. "Requirements for Grid Supporting Inverter in Relation with Frequency and Voltage Support" International Conference & Exhibition on Electricity Distribution: CIRED 2023. 2023.

[5] Zhang, Ziqian, et al. "Domain of attraction's estimation for grid connected converters with phase-locked loop." IEEE Transactions on Power Systems 37.2 (2021): 1351-1362.

[6] Zhang, Ziqian, et al. "Study of stability after low voltage ride-through caused by phase-locked loop of grid-side converter." International Journal of Electrical Power & Energy Systems 129 (2021): 106765.

[7] Brestan, Maximilian Heinz, et al. "Performance of a digital distance protection relay during short circuit in presence of a converter connected grid." International Conference & Exhibition on Electricity Distribution: CIRED 2023. 2023.

[8] Zhang, Ziqian, Lothar Fickert, and Yongming Zhang. "Power hardware-in-the-loop test for cyber physical renewable energy infeed: Retroactive effects and an optimized power Hardware-in-the-Loop interface algorithm." 2016 17th International Scientific Conference on Electric Power Engineering (EPE). IEEE, 2016.

[9] https://www.e-control.at/documents/1785851/1811582/TOR+Erzeuger+Typ+A+V1.0.pdf/6342d021-a5ce-3809-2ae5-28b78e26ff04d?t=1562757767659

[10] CIGRE, TF. "Benchmark systems for network integration of renewable and distributed energy resources." Benchmark systems for network integration of renewable and distributed energy resources", technical brochure, version 21 (2014).